\documentclass[11pt,twoside]{article}  
\usepackage{asp2006}
\usepackage{adassconf}
\usepackage{graphics}
\usepackage[leftcaption]{sidecap}
\usepackage{wrapfig}

\begin{document}   

%
%

\paperID{P6.6}

%

\title{Galaxy Structural Parameters in Source Extractor}

%
%
%
%
%

\markboth{B.W. Holwerda}{Galaxy Strucutral Parameters in Source Extractor}

%
%
%
%

\author{B. W. Holwerda}
\affil{Space Telescope Science Institute}

%

\contact{Benne W. Holwerda}
\email{holwerda@stsci.edu}

%
%
%

\paindex{Holwerda, B.W.}

%



\begin{abstract}          
Over the last decade, the Concentration, Asymmetry and Smoothness (CAS), as well as the M20 and GINI parameters have become popular to automatically classify distant galaxies in images. Ellipticals, spirals and irregular galaxies all appear to occupy different regions of this parameter space. At the same time, the Source Extractor (SE) program has become the mainstay to produce object catalogs from large image surveys.
A logical next step would be to incorporate the structural parameters into the Source Extractor software. There are however several problems that arise: 1) the CAS parameters are fits to the images and Source Extractor eschews fits in the interest of speed, 2) the definition of the structural parameters changed over time.
Now that there is a clear and agreed-upon definition of the structural parameters, I am incorporating computed versions in the Source Extractor code (v2.5). The fitted CAS parameters are available for the GOODS-N/S fields and I compare the computed structural parameters to those found by the previous fits. My goal is to expand the source structure information in Source Extractor catalogs in order to improve automatic identification of sources, specifically of distant galaxies. 

The computed parameters perform reasonably close to the fitted versions but noise appears in faint objects due to a lack of information. For a subset of objects, the asymmetry signal is outside the SE boundaries and Smoothness still fails to compute for many objects. Type classification based on the SE parameters still lacks resolving power.
\end{abstract}

%
%

\section{Galaxy Structural Parameters}

There are six parameters that are popular for galaxy classification: Concentration (C), Asymmetry (A), Smoothness/Clumpiness (S), Moment of the top 20 \% pixels (M20), the GINI-parameter (G) and the ellipticity of the object (E):

\noindent {\bf Concentration}:
%
\cite{Conselice03} defines Concentration as the $5 ~ \times ~ log_{10}(r_{80}/r_{20})$ with $r_{80}$ and $r_{20}$ the circular radii encompassing 80\% and 20\% of the flux are computed by Source Extractor (FLUX\_RADIUS with PHOT\_FLUXFRAC at 0.8 and 0.2).  Different flux percentages are sometimes used, e.g., the SDSS uses $r_{90}/r_{50}$. Since the computation of radii is done in SE, one would expect an straightforward implementation of the Concentration parameter in SE. 

\noindent {\bf Asymmetry}:
Asymmetry is the absolute difference of the original object (I) with the same object, rotated by $180^{o}$ (R), and divided by the total flux of the object: $A = abs(I-R)/I$. Asymmetry is normally fit to objects with the x and y values of the center of rotation as variables. The information --position and value of all the pixels belonging to an object-- is available in the SE data-structure. Thus, implementation of a calculated --not fitted-- version of asymmetry is possible.

\noindent {\bf Smoothness}:
Smoothness is similar to Asymmetry but one subtracts a smoothed version of the object from itself. This parameterÕs definition has changed the most over time with both shape and size of the smoothing kernel changing significantly. At present the smoothing is done with a boxcar smooth using 0.1 Petrosian Radius as the smoothing kernel. 

\noindent {\bf M20}:
\cite{Lotz04} introduced the second-order moment of the brightest 20\% of the galaxyÕs flux ($\rm M20 = log_{10} \left( {\Sigma M_i /M_{tot}} \right), ~  with ~  \Sigma f_i  <  0.2 ~  f_{tot}$) compared to the total second moment of the object ($\rm M_{tot} = \Sigma_i^n f_i [(x_i -x_c)^2 + (y_i -y_c)^2] $). SE already works with second order moments but the implementation would need an ordered list of pixel values ($f_i$) and positions ($x_i$ and $y_i$). 

\noindent {\bf GINI}:
The GINI parameter is the second parameter used by \cite{Lotz04} and  \cite{Abraham03}. It is the 
area between the distribution of pixel values and a uniform distribution of pixel values. 
\cite{Lotz04} present a computationally cheaper version which uses a ordered list of pixel values: 
$G = {1 \over \bar{f} n(n-1)} \Sigma_i^n (2i-n-1) f_i$. 

\noindent {\bf Ellipticity}:
Ellipticity is defined by the Source Extractor manual as: $E ~ = ~ 1  -  b/a$, 
where $a$ is the major and $b$ the minor axis of the object. SE computes the values of $a$ and $b$ from the second-order moments of the pixels. 

The above parameters, except Smoothness, are used in the Zurich Estimator of Structural Type \citep[ZEST,][]{Scarlata07}. They determined the principal components of the above parameter-space from SDSS and COSMOS data and published the eigenvectors. So given the above parameters, one could easily determine the ZEST galaxy type-classification. 

\section{Implementation and Test Data}

The above parameters are implemented in \htmladdnormallinkfoot{Source Extractor}{http://terapix.iap.fr/} v2.5 code \citep[SE][]{se, seman} with two general classes: object and FIELD versions. The object version is computed from the pixels that SE has assigned to the object. The FIELD version is computed within the 1.5 Petrosian radius aperture. 

The Great Observatories Origins Deep Survey (GOODS) provides us with 2 deep mosaics made with {\em HST/ACS} in several filters. CAS parameters were determined by Chris Conselice and collaborators \citep[see also][]{Bundy05}. 
The SE parameters and catalogs used by the GOODS team are also \htmladdnormallinkfoot{public}{http://archive.stsci.edu/pub/hlsp/goods/catalog_r1/}.
I re-ran SE with the same parameters as the GOODS team but now with added structural parameters on the central fields of GOODS-S (22, 23, 24, 32, 33 and 34 in z filter). A cross-correllation of the catalogs gives the control for SE implementation. The SE run took substantially longer with the structural parameters than without. FIELD parameters substantially slow down SE, object ones inflict no performance penalty.
\begin{figure}[t]
\epsscale{0.65}
\plotone{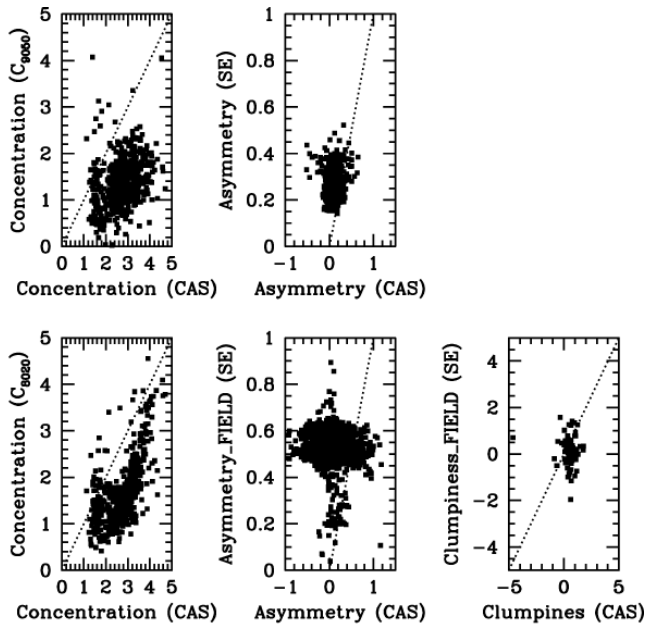}
\caption{\label{fig:P6.6:fig1} The relations between SE values and those determined by C. J. Conselice (CAS). {\bf left panels:} concentration indices from SE $C_{8020} = r_{80}/r_{20}$ and $C_{9050} = r_{90}/r_{50}$, and the concentration value provided by C. Conselice. The $C_{8020}$ still suffers from an offset. {\bf middle panels} asymmetry, both the object and computed with 1.5 Petrosian radii (FIELD), compared to the CAS value. The FIELD version can be computed for just any entry in the catalogs, resulting in a lot of noisy entries with A$\sim$0.5. {\bf right panel}: the smoothness parameter only resulted in overlap with the CAS catalog for the FIELD version. Smoothness fails repeatedly in either catalog, making it unreliable for now.}
\end{figure}
\section{Performance of Structural Parameters}

Figure \ref{fig:P6.6:fig1} shows the relation between the values from the CAS fits and the values determined by the SE on GOODS-S. The relation between concentration parameters indicate that the GOODS CAS catalog simply uses $C = r_{80}/r_{20}$.  The Asymmetry parameter becomes very much noisier in the FIELD version. Smoothness measures fail often for the original CAS fit (S=0) but also for the SE implementation. Only the FIELD version of Smoothness produces some overlap. 

\begin{figure}[t]
\centering
\epsscale{0.3}
\plotone{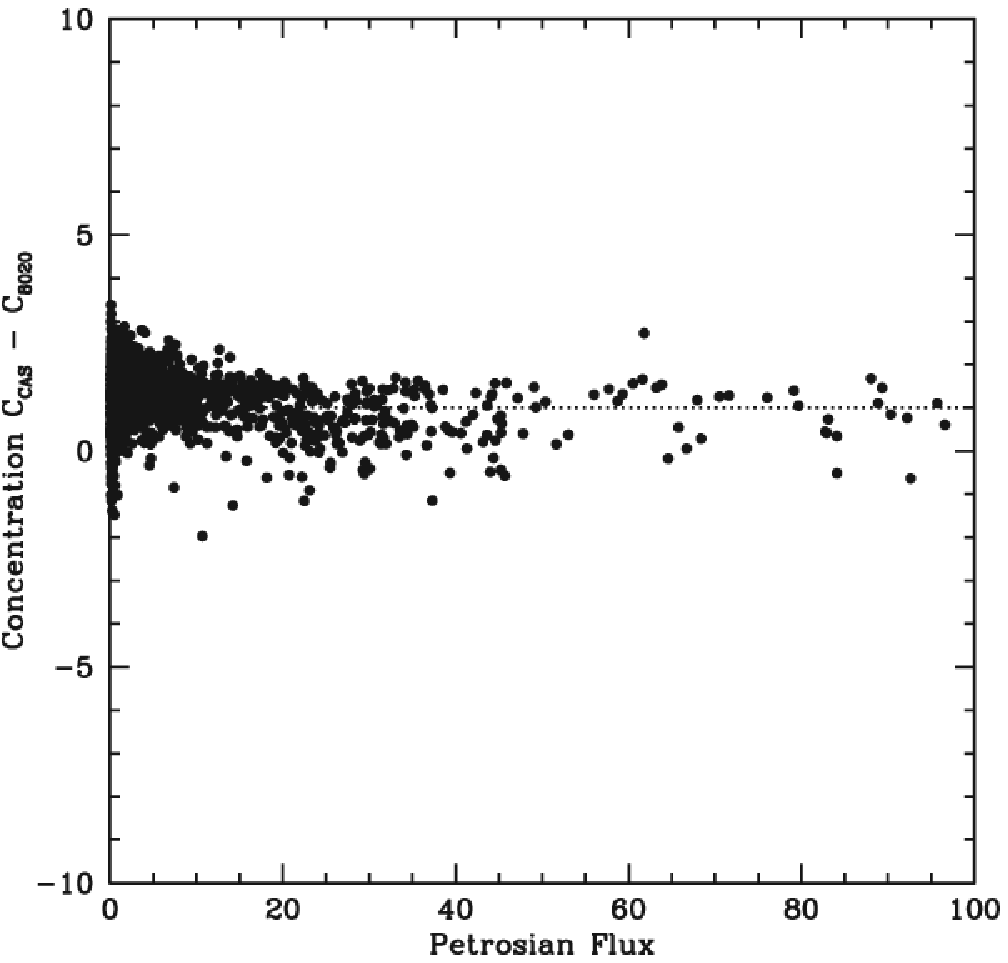}
\plotone{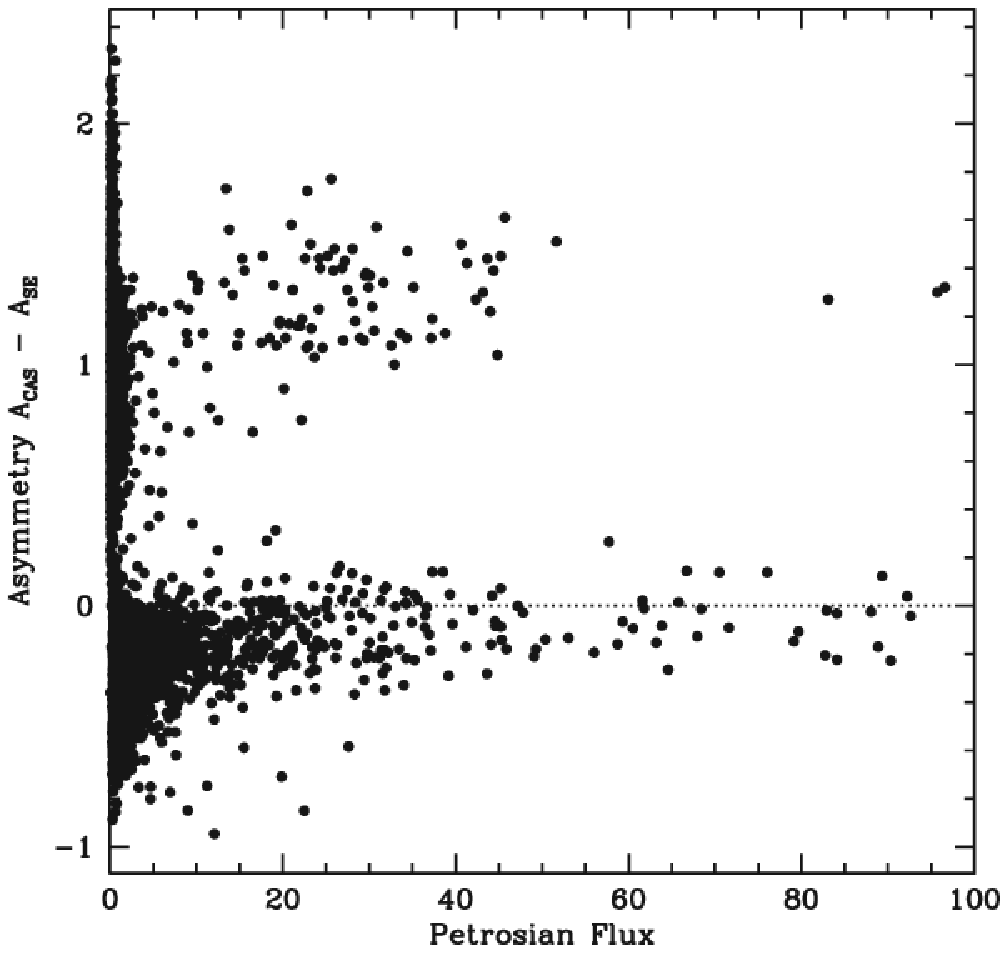}
\plotone{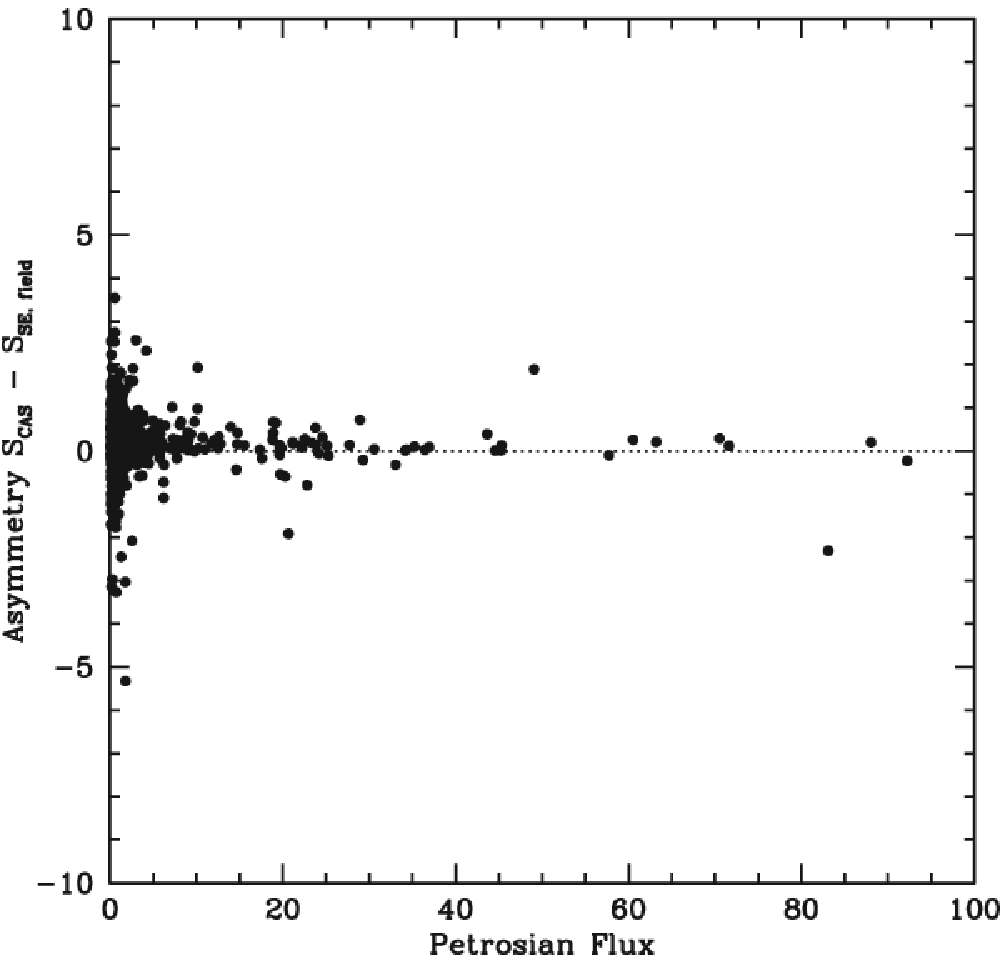}

\caption{\label{fig:P6.6:fig2}The difference between the CAS and the SE values as a function of flux. As expected, at lower flux, the parameters become erratic. Concentration suffers from an unexplained offset. Asymmetry is larger for CAS some cases because some information is outside the group of pixels assigned to the object by SE.  The Smoothness\_FIELD parameter works, but only for a fraction of the objects in the catalog.}
\end{figure}

Figure \ref{fig:P6.6:fig2} shows the difference between the SE determined value and those from the CAS catalog for Concentration, Asymmetry and Smoothness. Noise increases at lower fluxes, as can be expected with the dearth of information for faint objects. For some large objects, most of the Asymmetry information is {\em outside} the object as defined by SE.

ZEST Type classification based on structural types seem possible as well as an estimate of Sersic index (n) from the ratio of Petrosian and effective radius \citep[$r_{50}$,][]{Graham05}. However, the SE classification does not differentiate types well enough yet.

\section{Conclusions}

Based on the first application of the modified SE v2.5 on the GOODS-S fields, I conclude the following:
\begin{itemize}
\itemsep=0.2pt
\item Galaxy structural parameters can be implemented in Source Extractor, opening up possibilities for structural classification in large-scale surveys.
\item Concentration values suffer from an unexplained offset (Fig. \ref{fig:P6.6:fig1}).
\item Faint emission at the edge of objects may dominate Asymmetry (Fig. \ref{fig:P6.6:fig2}). 
\item Smoothness fails often in both the original catalog and the SE approximation(Fig. \ref{fig:P6.6:fig1}) but agrees when computed (Fig. \ref{fig:P6.6:fig2}).
\item M20 and GINI are clearly defined and appear to work.
\item ZEST galaxy type based on SE is not robust yet (all objects are t$\sim$2).
\end{itemize}
\acknowledgements 
The author would like to thank I. Smail, C.J. Conselice, J. Lotz and C. Scarlata for their code, catalogs,  and comments.


\begin{thebibliography}{}

\bibitem[\protect\astroncite{{Abraham} et~al.}{1997}]{Abraham96}
{Abraham}, R.~G., {van den Bergh}, S., {Glazebrook}, K., {Ellis}, R.~S.,
  {Santiago}, B.~X., {Surma}, P., and {Griffiths}, R.~E.: 1997,
\newblock {\em \apjs} {\bf 107}, 1+,

\bibitem[\protect\astroncite{{Abraham} et~al.}{2003}]{Abraham03}
{Abraham}, R.~G., {van den Bergh}, S., and {Nair}, P.: 2003,
\newblock {\em \apj} {\bf 588}, 218

\bibitem[\protect\astroncite{{Bertin} and {Arnouts}}{1996}]{se}
{Bertin}, E. and {Arnouts}, S.: 1996,
\newblock {\em \aaps} {\bf 117}, 393

\bibitem[\protect\astroncite{{Bundy} et~al.}{2005}]{Bundy05}
{Bundy}, K., {Ellis}, R.~S., and {Conselice}, C.~J.: 2005,
\newblock {\em \apj} {\bf 625}, 621

\bibitem[\protect\astroncite{{Conselice}}{2003}]{Conselice03}
{Conselice}, C.~J.: 2003,
\newblock {\em \apjs} {\bf 147}, 1

\bibitem[\protect\astroncite{{Graham} et~al.}{2005}]{Graham05}
{Graham}, A.~W., {Driver}, S.~P., {Petrosian}, V., {Conselice}, C.~J.,
  {Bershady}, M.~A., {Crawford}, S.~M., and {Goto}, T.: 2005,
\newblock {\em \aj} {\bf 130}, 1535

\bibitem[\protect\astroncite{{Holwerda}}{2005}]{seman}
{Holwerda}, B.~W.: 2005,
\newblock {\em astro-ph/0512139}

\bibitem[\protect\astroncite{{Lotz} et~al.}{2004}]{Lotz04}
{Lotz}, J.~M., {Primack}, J., and {Madau}, P.: 2004,
\newblock {\em \aj} {\bf 128}, 163

\bibitem[\protect\astroncite{{Scarlata} et~al.}{2007}]{Scarlata07}
{Scarlata}, C., {Carollo}, C.~M., {Lilly}, S., {Sargent}, M.~T., {Feldmann},
  R., {Kampczyk}, P., {Porciani}, C., {Koekemoer}, A., {Scoville}, N., {Kneib},
  J.-P., {Leauthaud}, A., {Massey}, R., {Rhodes}, J., {Tasca}, L., {Capak}, P.,
  {Maier}, C., {McCracken}, H.~J., {Mobasher}, B., {Renzini}, A., {Taniguchi},
  Y., {Thompson}, D., {Sheth}, K., {Ajiki}, M., {Aussel}, H., {Murayama}, T.,
  {Sanders}, D.~B., {Sasaki}, S., {Shioya}, Y., and {Takahashi}, M.: 2007,
\newblock {\em \apjs} {\bf 172}, 406

\end{thebibliography}
\end{document}